\documentclass[12pt]{article} 

\usepackage{geometry}
\usepackage[english]{babel}
\usepackage[utf8]{inputenc}
\usepackage[T1]{fontenc}
\usepackage{algorithm}
\usepackage[noend]{algpseudocode}
\usepackage{amsmath}
\usepackage{amssymb}
\usepackage{graphicx}
\usepackage{amsthm}
\usepackage{amsfonts}
\usepackage{subcaption}
\usepackage{dsfont}
\usepackage[round,sort,comma,authoryear]{natbib}

\usepackage{verbatim}
\usepackage{latexsym}
\usepackage{graphics}
\usepackage{epsfig} 
\usepackage{fancyhdr} 
\usepackage{setspace} 
\usepackage[colorinlistoftodos]{todonotes}
\usepackage[colorlinks=true, allcolors=blue]{hyperref}

\renewenvironment{quotation}{%
   \list{}{%
     \leftmargin0.5cm   
     \rightmargin\leftmargin
   }
   \item\relax
}
{\endlist}

\def\BState{\State\hskip-\ALG@thistlm}
\usepackage{float}
\newfloat{algorithm}{t}{lop}

\usepackage{graphicx}
\usepackage{multirow}

\begin{document}

\title{Auditable Blockchain Randomization Tool} 
		  
\author{
	Olivia Saa\footnote{IME-USP -- Institute of Mathematics and Statistics  of the University of S\~{a}o
		Paulo, Rua do Mat\~{a}o 1010, 05508-090, S\~{a}o Paulo, Brazil. \  e-mails: 
		\texttt{olivia@ime.usp.br} and \texttt{olivia.saa@iota.org} } \ \ \ \  
	Julio Michael Stern\footnote{   
 IME-USP -- Institute of Mathematics and Statistics 
of the University of S\~{a}o Paulo.  
 Rua do Mat\~{a}o 1010, 05508-090, S\~{a}o Paulo, Brazil. \ 
  e-mails: \texttt{jstern@ime.usp.br} and \texttt{jmstern@hotmail.com} 
  }
    }

\lhead{O. Saa, J.M. Stern}
\rhead{Auditable Blockchain Randomization Tool} 
    
   	
\maketitle 
   
\begin{abstract} 
Randomization is an integral part of well-designed statistical trials, and is also a required procedure in legal systems,  see \citet{Marcondes2019}.   
This paper presents an easy to implement randomization protocol that assures, in a formal mathematical setting, a statistically sound,  computationally efficient, cryptographycally secure, traceable and auditable randomization procedure that is also resistant to collusion and manipulation by participating agents.  
 \vspace{0mm}  
\end{abstract} 

 \begin{flushright} 

 \noindent 
 \textit{Meos tam suspicione quam crimine iudico carere oportere}.  \\  
 My people should be free from either crime or suspicion. \\  
 Julius Caesar (62BC), 
 in Suetonius (119CE, Sec.I.74.2). 
 \vspace{2mm} 

 \end{flushright}

\section*{Randomization: Bad and Good Practices} 

Randomization is a technique used in the design of statistical experiments: in a clinical trial, for example, patients are randomly assigned to distinct groups receiving different treatments with the goal of studding and contrasting their effects. 
Randomization is nowadays considered a golden standard in statistical practice; its motivation is to prevent systematic biases (like an unfair or tendentious assignment process) that could distort (unintentionally or purposely) the conclusions of the study.   
For further comments on randomization see \citet{Pearl2000, Pearl2004} and \citet{Stern2008}, for Bayesian perspectives see \citet{Basu1988} and \citet{Gelman2003}.        
In the legal context, randomization (also known as sortition or allotment) is routinely used for the selection of jurors or judges assigned to a given judicial case; see \citet{Marcondes2019}.  
   
Rerandomization is the practice of rejecting and discarding (for whatever reason) a given randomized outcome, that is subsequently replaced by a new randomization.  
Repeated rerandomization can be used to completely circumvent the haphazard, unpredictable or aimless nature of randomization, allowing a premeditated selection of a final outcome of choice. 
There are advanced statistical techniques capable of blending the best characteristics of random and intentional sampling, see for example \citet{Fossaluza2015}, \citet{Lauretto2012, Lauretto2017}, and \citet{MorganRubin2012, MorganRubin2015}.       
Nevertheless, rerandomization is often naively used, or abused, with the excuse of (subjectively) ``avoiding outcomes that do not look random enough'', see for example 
\citet{Bruhn2009} and \citet{Ruxton2006}. 
In the legal context, spurious manipulations of the randomization process are often linked to  fraud, corruption and similar maladies, see \citet{Marcondes2019} and references therein. 

In order to comply with the best practices for randomization processes, \citet{Marcondes2019} recommends the use of computer software having a long list of characteristics, for example, being efficient and fully auditable, well-defined and understandable, sound and flexible, secure and transparent. Such requirements are expressed by the following (revised) \textit{desiderata for randomization procedures}:

\begin{quotation} 

\textit{Given the juridical and social importance of the themata under scrutiny, we believe that it is important to develop randomization procedures in full compliance with the following desiderata: 
(a) Statistical soundness and computational efficiency, see \citet{Hammersley1964}, \citet{Haramoto2008}, \citet{Knuth1997}, and \citet{Ripley1987}; 
(b) Procedural, cryptographical and computational security, see \citet{Boyar1989}, \citet{lecuyer2012}, \citet{Aumasson2017} and \citet{katz2014}; 
(c) Complete auditability and traceability, see \citet{Haber1991}, \citet{nakamoto2008}; and \citet{Wattenhofer2017}; 
(d) Any attempt by participating parties or coalitions to spuriously influence the procedure should be either unsuccessful or be detected,  see \citet{Goldschlag1998}; 
(e) Open-source programming; 
(f) Multiple hardware platform and operating system implementation; 
(g) User friendliness and transparency, see \citet{Parikh2012} and \citet{Stern2018}; 
(h) Flexibility and adaptability for the needs and requirements of multiple application areas (like, for example, clinical trials, selection of jury or judges in legal proceedings, and draft lotteries), see \citet{Marcondes2019}. } 	
	
\end{quotation} 	 

Such requirements conflate several complementary characteristics that may seem, at first glance, incompatible.    
For example, strong security is often (but wrongly) associated with excessive secrecy, a doctrine known as ``security by obscurity'', computer routines may be efficient but are often tough as hard to audit, and mathematically well-defined algorithms may be perceived as hard to understand. 
The bibliographical references given in the formerly stated \textit{desiderata for randomization procedures} already hint at  technologies that can be used to achieve a fully compliant randomization procedure, most preeminently, the blockchain. 
This is the key technology supporting modern public ledgers, cryptocurencies, and a host of related applications.

A technical challenge for the application under scrutiny is the generation of pseudo-random number sequences that reconcile complementary properties related to computational efficiency, statistical soundness, and cryptographic security. 
In this respect, the excellent statistical and computational characteristics of modern linear recurrence pseudo-random number generators, like \citet{Haramoto2008}, can be reconciled with the needs concerning unpredictability and cryptographic security by appropriate starts and restarts of the linear recurrence generator.  
A sequence start for a linear recurrence generator is defined by a \textit{seed} specified by a vector of (typically 1 to 64) integers, while a restart is defined by a \textit{jump-ahead} or \textit{skip-ahead}  specified by a single  integer (kept small relative to the generator's full period), see \citet{lecuyer2012}.  

Unpredictable and cryptographically secure  
\textit{seeds} and \textit{jump-aheads} can be provided by high entropy  bit streams extracted from blockchain transactions, an idea that has already been explored in the works of \cite{Bonneau2015} and \citet{Popov2017}. 

The next section develops a possible implementation of a fully compliant core randomization protocol based on blockchain technology, and also makes a simple prototype available for study and further research\footnote{www link to be added at publication time}. 
Moreover, in order to make it simple and easy to use, we  develop the prototype on top of a readily available crypto-currency platform. We use Bitcoin for this example, but other alternatives like Ethereum or other cryptocurrencies whose miners work under the same incentives model can be used with minor adaptations.

\section*{Core Randomization Protocol in Blockchain}

We intend to establish a protocol able to deliver on demand pseudo random numbers, from a auditable and immutable ledger. The procedure will start as follows: the user (the part that wants to receive a random number) shall send a Bitcoin transaction with a register of its purpose embedded\footnote{One way to embed a message in a transaction is using the OP$\_$RETURN script, which allows to store up to 40 bytes in a transaction} in it. 
The recipient of this transaction may be a proxy representing a competent authority, a pertinent regulatory agency, an agreed custodian, etc.  
When this\footnote{If someone tries to generate more than one transaction for a same purpose, just take the one that was attached first.} transaction is first attached to the blockchain, we concatenate the transaction ID (a 32 bytes, hexadecimal number) and the block header (a 80 bytes, hexadecimal number). This resulting 112 bytes hexadecimal number will be the input for some known Verifiable Delay Function (VDF), that should be calibrated accordingly to the purpose of the random number. For instance, a less critical purpose should have a VDF that delays the result in just a few seconds, or even skip completely the VDF step. A critical purpose, with significant interests involved, should have a more complex VDF, with a delay of minutes or even hours. The final result, after the VDF, will be the source for our seeds and jump-aheads.

With the aid of this protocol, one is able to find a different pseudo-random number for each user that demands it. Note that the user does not have any incentive to try to modify its transaction ID, because he does not have any control of the block header. We assume that the user and the miner are not the same person, so a miner will only be interested in trying to control his block header if he is paid to do so. Since the last stage of our protocol involves the calculation of a VDF, it will take a certain amount of time to the miner to decide if the the block he has found will be of interest of the user. Thus, he might even lose his block, if some other miner broadcasts a block of his own before he finishes calculating the VDF. 

In the following subsection, the miner's payoff and the necessary delay $T$ for the Verifiable Delay Functions will be explicitly calculated. 

\subsubsection*{Preventing Collusion for Spurious Manipulation} 

Suppose a malicious user tries to bribe a miner that controls a fraction $p$ of the network's computational power. A prize $P=nB$, where $B$ is the Bitcoin block reward, will be paid to the miner if he successfully mines what we call a "desirable block": a block that will deliver a random number in a set $A$, chosen by the malicious user. Let also $\lambda$ be the average rate of incoming blocks and $q$ the probability of a randomly generated number being an element of $A$, i.e., the measure of the set of desirable results for the malicious user. Finally, let $T$ be the expected amount of time needed for the VDF calculations. The moment a miner finds a block that can be accepted by the network, he faces the decision of broadcasting it before checking the VDF, or calculating the VDF before broadcasting. If he decides to check the VDF before broadcasting, he might start another attempt to find a block rightaway.

First, we calculate the expected absolute payoff for the first and second options, called $\mathds{E}_1$ and $\mathds{E}_2$, respectively. $\mathds{E}_1$ will be larger than $B$, since the miner might issue a desirable block by chance:

\begin{equation}
  \mathds{E}_1=  B+qP  = B(1+nq)
\end{equation}

On the other hand, if the miner chooses to calculate the VDF, he will receive the block reward and the prize $P$, but with a probability given by

\begin{align}
  \mathds{E}_2=&  (B+P)q\mathds{P} \{ \text{no other node finding a block before }t=T \} \\
   &+  (B+P)(1-q)\mathds{P}\{\text{successfully mining a desirable block in another attempt} \} \nonumber\\
    =& B(1+n)q\exp(-(1-p)\lambda T)\nonumber\\
 &+ B(1+n)(1-q)\sum_{i=1}^{\infty}\mathds{P}
\{\text{successfully mining a desirable block after }i\text{ attempts}\}\nonumber
\end{align}

The probabilities inside the summation, in the last equation, can be calculated as the product of the probability of finding a desirable block after $i$ attempts (that will be a geometric distribution with probability of success $q$) and the probability of finding and checking $i$ blocks before the rest of the network mines one \footnote{
   $P\{\text{attacker finding and analyzing }i\text{ blocks before another node mining one}\}$ \\
   $=\displaystyle\int_{t=0}^{\infty} p\lambda\exp(-p\lambda t)\frac{(p\lambda t)^{i-1}}{(i-1)!} \exp(-(1-p)\lambda (t+T)) dt$ 
   $=p^i\exp(-(1-p)\lambda T)$
    },  
resulting:


\begin{align}
\mathds{E}_2=&B(1+n)\left[q\exp(-(1-p)\lambda T)+(1-q)\sum_{i=1}^{\infty} q(1-q)^{i-1} p^i\exp(-(1-p)\lambda T)\right]\nonumber\\
 =&B(1+n)\exp(-(1-p)\lambda T)\left(q+\frac{(1-q)pq}{1-p+pq}\right)
\end{align}

Finally, in order to make accepting the bribe not lucrative, we must have $\mathds{E}_1>\mathds{E}_2$, i.e.:
\begin{equation}
\lambda T>\frac{1}{1-p}\log\left(\frac{1+n}{1+nq}\frac{q}{1-p+pq}\right)
\end{equation}

Since for every $n>0$ we have $\frac{1+n}{1+nq}<\frac{1}{q}$, if we choose $\lambda T^{*} = \frac{1}{1-p}\log\left(\frac{1}{q}\frac{q}{1-p+pq}\right) $, we guarantee that the attack will not be lucrative for any bribe $P=nB$. Also, since it can be assumed that $p<1/2$, a value $\lambda T^{*}=2\log\left(\frac{2}{1+q}\right)<2\log(2)$ will be high enough to prevent an attack for any bribe and any acceptable value of $p$.

 \section*{Conclusions and Final Remarks} 
  
We formalized a simple and effective protocol to generate on demand pseudo random numbers, in a fully auditable way. We have demonstrated that none of the involved parts has enough financial incentives to try to affect the random number outcome: the part that issues the transaction lacks this power, since it does not have any control on the block header; and the miners do not have enough financial incentives to collude with an attacker, provided a suitable Verifiable Delay Function is applied. 
   
The essentially decentralized, yet completely traceable and auditable nature of the protocol presented in this article, makes the resulting randomization process eminently reliable without recourse of 
blind trust in any central authority. 
The authors believe the adoption of such a protocol by the the Brazilian Supreme Court (STF), as recommended in \citet{Marcondes2019}, would significantly  increase public confidence in the judicial system and be a contributing factor for political and social stability.  

\subsubsection*{Acknowledgments} 

 The authors are grateful for the support received from 
 IME-USP -- the Institute of Mathematics and Statistics of the University of S\~{a}o Paulo and for the advice of Prof. 
 Serguei Popov. 
 The authors also received support from   
 FAPESP -- the State of S\~{a}o Paulo Research Foundation (grants CEPID-CeMEAI 2013/07375-0 and CEPID-Shell-RCGI 2014/50279-4); 
 CNPq -- the Brazilian National Counsel of Technological and Scientific Development (grant PQ 301206/2011-2 and GD 140490/2016-7);  
 ABJ -- the Brazilian Jurimetrics Association;  
 STF -- Supremo Tribunal Federal (the Brazilian Supreme Court), which motivated the study and provided the data analysed in \citet{Marcondes2019}; the IOTA Foundation; and received helpful comments and advice from Adilson Simonis, \'{A}lvaro Machado Dias, Julio Trecenti, Rafael Bassi Stern and Marcelo Guedes Nunes. 
 
 \bibliographystyle{plainnat}

\begin{thebibliography}{99}


\bibitem[Aumasson (2017)]{Aumasson2017} Aumasson, Jean-Philippe (2017). 
Serious Cryptography: A Practical Introduction to Modern Encryption. No Starch Press. 
 
\bibitem[Basu (1988)]{Basu1988} Basu, D.; Ghosh, J.K. (ed.) (1988), Statistical Information and Likelihood, A Collection of Essays by Dr.Debabrata Basu, Lecture Notes in Statistics, 45, Springer.

\bibitem[Boneh et al. (2018)]{Boneh2018}  Boneh, D.; Bonneau, J.; Bünz, B.; Fisch, B. Verifiable delay functions. Cryptology ePrint Archive, Report 2018/601, 2018. 
\texttt{https://eprint.iacr.org/2018/601}

\bibitem[Bonneau et al. (2015)]{Bonneau2015} Bonneau, Joseph, Jeremy Clark, and Steven Goldfeder. On Bitcoin as a public randomness source. IACR Cryptology ePrint Archive 2015 (2015): 1015. 


\bibitem[Boyar (1989)]{Boyar1989} Boyar, J. (1989). Inferring Sequences Produced by Pseudo-Random Number Generators. Journal of the ACM, 36, 1, 129-141. 


\bibitem[Bruhn and McKenzie (2009)]{Bruhn2009} Bruhn, M.; McKenzie, D. (2009). In Pursuit of Balance: Randomization in Practice in Development Field Experiments. American Economic Journal, Applied economics, 1, 4, 200-232.


\bibitem[Fossaluza (2015)]{Fossaluza2015} Fossaluza, V.; Lauretto, M.S.; Pereira, C.A.B.; Stern, J.M. (2015). Combining Optimization and Randomization Approaches for the Design of Clinical Trials. 
Springer Proceedings in Mathematics and Statistics, Vol. 118, Ch. 14, p. 173-184.  

\bibitem[Gelman et al. (2003)]{Gelman2003} Gelman, A.; Carlin, J.B.; Stern, H.S.; Rubin, D.B. (2003). Bayesian Data Analysis, 2nd ed. NY: Chapman and
Hall / CRC 

\bibitem[Goldschlag and Stubblebine (1998)]{Goldschlag1998} Goldschlag, D. M.; Stubblebine S. G. (1998). Publicly Verifiable Lotteries: Applications of Delaying Functions. p.214-226 in International Conference on Financial Cryptography. Berlin: Springer. 
  
\bibitem[Haber and Stornetta (1991)]{Haber1991} Haber, S., Stornetta, W. (1991). How to time-stamp a digital document. Journal of Cryptology. 3 (2): 99–111. 

\bibitem[Hammersley and Handscomb (1964, ch.3)]{Hammersley1964} Hammersley, J.M., Handscomb, D.C.  (1964). Monte Carlo Methods. London: Chapman and Hall. 

\bibitem[Haramoto et al. (2008)]{Haramoto2008} Haramoto, H., M., Matsumoto, T., Nishimura, F., Panneton, and P. L’Ecuyer. 2008. Efficient Jump Ahead for F2-Linear Random Number Generators. INFORMS Journal on Computing, 20 (3): 290-298. 

\bibitem[Katz and Lindell (2014)]{katz2014}  Katz, Jonathan; Lindell, Yehuda (2014).
\textit{Introduction to Modern Cryptography}.   Chapman and Hall. 

\bibitem[Kelsey et al. (1997)]{Kelsey1997} Kelsey, John, et al. (1997). Secure applications of low-entropy keys. International Workshop on Information Security. Berlin: Springer. 
  
\bibitem[Knuth (1997, ch.3)]{Knuth1997} Knuth, D. E. (1997). The Art of Computer Programming, vol. 2: Seminumerical Algorithms. 3rd ed. Addison-Wesley Longman Publ. Co., Inc., Reading, MA.

\bibitem[Lauretto et al. (2012)]{Lauretto2012} Lauretto, M.S.; Nakano, F.; Pereira, C.A.B.; Stern, J.M. (2012). Intentional Sampling by Goal Optimization with Decoupling by Stochastic Perturbation. American Institute of Physics Conference Proceedings, 1490, 189-201. 
   
\bibitem[Lauretto et al. (2017)]{Lauretto2017} Lauretto, M.S.; Stern, R.B.; Morgan, K.L.;  Clark, M.H.; Stern, J.M. (2017). Haphazard Intentional Sampling and Censored Random Sampling to Improve Covariate Balance in Experiments.  American Institute of Physics Conference Proceedings, 1853, 050003, 1-8.  

\bibitem[L'Ecuyer (2012)]{lecuyer2012} L'Ecuyer, P. (2012). Random number generation. pp.35-71 in Handbook of Computational Statistics. Springer, Berlin, Heidelberg. 

\bibitem[Lindley (1982)]{Lindley1982} Lindley, D.V. The Role of Randomization in Inference. In PSA: Proceedings of the Biennial meeting of the philosophy of science association, pages 431–446. JSTOR, 1982.

\bibitem[Marcondes et al.(2019)]{Marcondes2019} 
Marcondes, D.; Peixoto C.; and Stern, J.M. Assessing Randomness in Case Assignment: The Case Study of the Brazilian Supreme Court.  
\textit{Law, Probability and Risk},  
\texttt{https://doi.org/10.1093/lpr/mgz006} , 2019

\bibitem[Morgan and Rubin (2012)]{MorganRubin2012} Morgan, K.L. and Rubin,D.B. Rerandomization to improve covariate balance in experiments. The Annals  
of Statistics, 40(2):1263–1282, 2012.

\bibitem[Morgan and Rubin (2015)]{MorganRubin2015} Morgan, K.L. and Rubin,D.B. Rerandomization to balance tiers of covariates. Journal of the American
Statistical Association, 110(512):1412–1421, 2015.

\bibitem[Nakamoto (2008)]{nakamoto2008} Nakamoto, Satoshi (2008). Bitcoin: A Peer-to-Peer Electronic Cash System.  \\   \texttt{https://bitcoin.org/bitcoin.pdf}  
 
\bibitem[Parikh and Pauly (2012)]{Parikh2012} Parikh, R.; Pauly, M. (2012). What Is Social Software? pp.3-13 in J.van Eijck et al. (2012). Games, Actions, and Social Software. LNCS 7010, Berlin: Springer-Verlag. 

\bibitem[Pearl (2000)]{Pearl2000} Pearl, J. Causality: models, reasoning, and inference. Cambridge University Press, 2000.

\bibitem[Pearl (2004)]{Pearl2004} Pearl, J. (2004). Simpson's paradox: An anatomy. Tech.Rep., Univ. of California, Los Angeles.  


 \bibitem[Pierrot (2016)]{Pierrot2016} Pierrot, C.; Wesolowski, B. (2016) Malleability of the blockchain's entropy. \\ 
 \texttt{https://eprint.iacr.org/2016/370.pdf} 

\bibitem[Popov (2017)]{Popov2017} Popov, Serguei. On a decentralized trustless pseudo-random number generation algorithm.  Journal of Mathematical Cryptology 11.1 (2017): 37-43.
  
\bibitem[Rabin (1983)]{Rabin1983} Rabin, Michael O. (1983). Transaction protection by beacons. Journal of Computer and System Sciences 27, 2, 256-267.
 
\bibitem[Ripley (1987, ch.2)]{Ripley1987} Ripley, B.D. (1987). Stochastic Simulation. NY: Wiley.

\bibitem[Ruxton and Colegrave (2006, Sec.3.4.6)]{Ruxton2006}  Ruxton, Graeme D.; Colegrave, Nick  (2006).
Experimental Design for the Life Sciences, 2nd ed. 
Sec. 3.4.6 Random samples and Representative Samples, p.59-60.  

\bibitem[Stern (2008)]{Stern2008} Stern, Julio Michael (2008). Decoupling, Sparsity, 
Randomization, and Objective Bayesian Inference. 
\textit{Cybernetics \& Human Knowing}, 15, 2, 49-68.

\bibitem[Stern (2018)]{Stern2018} Stern, J. M. (2018) Verstehen (causal/ interpretative understanding), Erkl\"{a}ren (law-governed description/ prediction),
and Empirical Legal Studies. Journal of Institutional and Theoretical Economics (JITE), 174, 105–114.  

\bibitem[Suetonius Tranquillus (119CE, 1979)]{Suetonius} Suetonius Tranquillus (119CE, 1979). The Lives of the Caesars, Vol.1. Harvard University Press. Sec.74.2. 
  
 
\bibitem[Wattenhofer (2017)]{Wattenhofer2017} Wattenhofer, R. (2017). Distributed Ledger Technology: The Science of Blockchain. Inverted Forest. 
   
\end{thebibliography}

\end{document}